\documentclass[prc,superscriptaddress,noshowpacs,unsortedaddress,twocolumn,showpacs,preprintnumbers,amsmath,amssymb]{revtex4-1}

\usepackage[dvipdfmx]{graphicx}
\usepackage{amsmath,amssymb,times}
\usepackage{color}
\usepackage{ulem}
\usepackage{bm}
\usepackage{here}

\def\lsim{~\,\makebox(1,1){$\stackrel{<}{\widetilde{}}$}\,~}

\newcommand{\beq}{\begin{equation}}
\newcommand{\eeq}{\end{equation}}
\newcommand{\bea}{\begin{eqnarray}}
\newcommand{\eea}{\end{eqnarray}}

\newcommand{\bfi}[1]{\mbox{\boldmath $#1$}}

\newcommand{\vK}{{\bfi K}}

\newcommand{\vs}{{\bfi s}}

\newcommand{\vrr}{{\bfi r}}
\newcommand{\vR}{{\bfi R}}

\def\a{\alpha}

\begin{document}
\title{Neutron-skin values and matter and neutron radii determined  \\
from reaction cross sections of proton scattering on $^{12}$C, $^{40,48}$Ca, $^{58}$Ni,  $^{208}$Pb
}

\author{Tomotsugu~Wakasa}
\affiliation{Department of Physics, Kyushu University, Fukuoka 819-0395, Japan}

\author{Shingo~Tagami}
\affiliation{Department of Physics, Kyushu University, Fukuoka 819-0395, Japan}

\author{Jun~Matsui}
\affiliation{Department of Physics, Kyushu University, Fukuoka 819-0395, Japan}

\author{Maya~Takechi}
\affiliation{Niigata University, Niigata 950-2181, Japan}

\author{Masanobu Yahiro}
\email[]{orion093g@gmail.com}
\affiliation{Department of Physics, Kyushu University, Fukuoka 819-0395, Japan}             

\date{\today}

\begin{abstract}
\begin{description}
\item[Background]
Very lately, the PREX and the CREX collaboration present skin values, 
$r_{\rm skin}^{208}({\rm newPREX2}) =0.278 \pm 0.078\ {\rm (exp)} \pm 0.012\ {\rm (theor.)}\,{\rm fm}$ and 
$r_{\rm skin}^{48}=0.121 \pm 0.026\ {\rm (exp)} \pm 0.024\ {\rm (model)}$, respectively.
We recently determined  a neutron-skin value  
$r_{\rm skin}^{208}=0.278 \pm 0.035$fm from measured reaction cross sections $\sigma_{\rm R}({\rm exp})$ 
of p+$^{208}$Pb scattering in a range of incident energies $10 \lsim E_{\rm in} \lsim 100$ MeV where 
the chiral  (Kyushu) $g$-matrix folding model is reliable for $^{12}$C+$^{12}$C scattering. 
The data $\sigma_{\rm R}({\rm exp})$ are available for proton scattering on $^{58}$Ni, $^{40,48}$Ca, $^{12}$C targets. 
\item[Purpose]
Our first aim is to test the Kyushu $g$-matrix folding model for p+$^{208}$Pb scattering  in 
$20 \lsim E_{\rm in} \lsim 180$~MeV. 
Our second aim is to determine skin values $r_{\rm skin}$ 
and matter and neutron radii, $r_{\rm m}$ and $r_{\rm n}$, for 
$^{208}$Pb, $^{58}$Ni, $^{40,48}$Ca, $^{12}$C from the $\sigma_{\rm R}({\rm exp})$. 
\item[Methods]
Our method is the Kyushu $g$-matrix folding model 
with the  densities scaled from the D1S-GHFB+AMP densities, where D1S-GHFB+AMP 
stands for  Gogny-D1S HFB (GHFB) with the angular momentum projection (AMP). 
\item[Results]
As for proton scattering, we find that our model is reliable  in $20  \lsim E_{\rm in} \lsim 180$ MeV. 
For $^{208}$Pb, the skin value deduced from $\sigma_{\rm R}({\rm exp})$ 
in $20  \lsim E_{\rm in} \lsim 180$ MeV is $r_{\rm skin}^{208}(\sigma_{\rm R})=0.299 \pm 0.020$~fm. 
Our results on $r_{\rm skin}$ are compared with  the previous works.
\item[Conclusion]
Our result $r_{\rm skin}^{208}(\sigma_{\rm R}) = 0.299 \pm 0.020$~fm
agrees with $r_{\rm skin}^{208}({\rm PREX2}) = 0.283\pm 0.071$~fm. In addition,  
our result  $r_{\rm skin}^{48}=0.103 \pm 0.022$~fm is consistent with the CREX value.
\end{description}
\end{abstract}

\maketitle


\section{Introduction}
\label{Introduction}

Many theoretical  predictions on the symmetry energy $S_{\rm sym}(\rho)$ have been made so far  
by taking several experimental and observational constraints on $S_{\rm sym}(\rho)$ and their combinations. 
In neutron star (NS), the $S_{\rm sym}(\rho)$ and its density ($\rho$) dependence influence strongly the nature within the star.
The symmetry energy $S_{\rm sym}(\rho)$ cannot be measured by experiment directly. 
In place of  $S_{\rm sym}(\rho)$, the neutron-skin thickness $r_{\rm skin}$ is measured to determine 
the slope parameter $L$, since a strong correlation between $r_{\rm skin}^{208}$ and $L$ is well known~\cite{RocaMaza:2011pm}.

Horowitz {\it et al.} \cite{PRC.63.025501} proposed a direct measurement 
for neutron skin thickness $r_{\rm skin}$ = $r_{\rm n} - r_{\rm p}$, where 
$r_n$ and $r_{\rm p}$ are the root-mean-square radii of neutrons  
and protons, respectively.

The PREX collaboration has reported a new value, 
\begin{equation}
r_{\rm skin}^{208}({\rm PREX2}) = 0.283\pm 0.071\,{\rm fm},
\label{eq:PREX2}
\end{equation}
combining the original Lead Radius EXperiment (PREX)  result \cite{PRL.108.112502,PRC.85.032501} 
with the updated PREX2 result \cite{Adhikari:2021phr}. 
Very lately, the PREX collaboration has presented an accurate value\cite{PREX:2021umo}, 
\begin{equation}
r_{\rm skin}^{208}({\rm newPREX2}) =0.278 \pm 0.078\ {\rm (exp)} \pm 0.012\ {\rm (theor.)}\,{\rm fm},
\label{eq:PREX2-v2}
\end{equation}
The  value is most reliable for $^{208}$Pb. The $r_{\rm skin}^{208}({\rm PREX2})$ value 
is considerably larger than the other experimental 
values that are significantly model dependent 
\cite{PRL.87.082501,PRC.82.044611,PRL.107.062502,%
PRL.112.242502}.
 As an exceptional case, a nonlocal dispersive-optical-model 
(DOM) analysis of ${}^{208}{\rm Pb}$ deduces 
$r_{\rm skin}^{\rm DOM} =0.25 \pm 0.05$ fm \cite{PRC.101.044303}
consistent with $r_{\rm skin}^{208}({\rm PREX2})$.

 Very recently, the CREX group has presented~\cite{CREX:2022kgg} 
\bea
r_{\rm skin}^{48}({\rm CREX})=0.121 \pm 0.026\ {\rm (exp)} \pm 0.024\ {\rm (model)}\,{\rm fm}. 
 \label{CREX-value}
 \eea
The CREX value  is most reliable for $^{48}$Ca.

 The $r_{\rm skin}^{208}({\rm PREX2})$ provides crucial tests for the equation of state (EoS) 
of nuclear matter~\cite{PRC.102.051303,AJ.891.148,AP.411.167992,EPJA.56.63,JPG.46.093003}.
For example, Reed {\it et al.} \cite{Reed:2021nqk} 
report a value of the slope parameter $L$ 
and examine the impact of such a stiff symmetry energy 
on some critical NS observables. 
They deduce 
\bea
L = 106 \pm 37=69	\text{--}  143 ~{\rm MeV}
\eea
from  $r_{\rm skin}^{208}({\rm PREX2})$.

\begin{figure}[H]
\begin{center}
 \includegraphics[width=0.45\textwidth,clip]{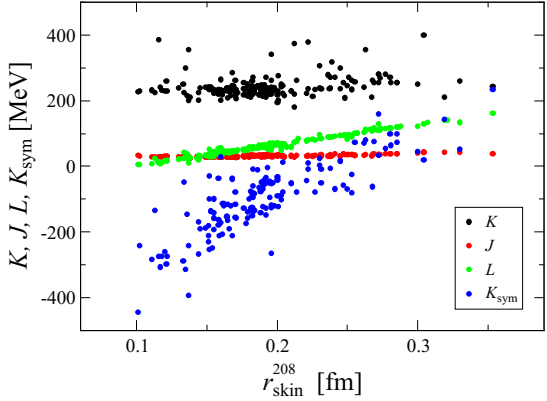}
  \caption{ 
 $r_{\rm skin}^{208}$ dependence of $J$, $L$, $K_{\rm sym}$, 
 where  $J$, $L$, $K_{\rm sym}$ are a constant term, the first-derivative and 
 the second-derivative term of the symmetry energy. 
 The dots show 206 EoSs taken from Table of Ref.~\cite{TAGAMI2022105155}. 
 Obviously, the correlation between $r_{\rm skin}^{208}$ and $L$ is linear. 
     }
 \label{Fig-L-skin}
\end{center}
\end{figure}

In Ref.~\cite{TAGAMI2022105155}, 
we  accumulated the 206 EoSs from Refs.~\cite{Akmal:1998cf,RocaMaza:2011pm,Ishizuka:2014jsa,Gonzalez-Boquera:2017rzy,D1P-1999,Gonzalez-Boquera:2017uep,Oertel:2016bki,Piekarewicz:2007dx,Lim:2013tqa,Sellahewa:2014nia,Inakura:2015cla,Fattoyev:2013yaa,Steiner:2004fi,Centelles:2010qh,Dutra:2012mb,Brown:2013pwa,Brown:2000pd,Reinhard:2016sce,Tsang:2019ymt,Ducoin:2010as,Fortin:2016hny,Chen:2010qx,Zhao:2016ujh,Zhang:2017hvh,Wang:2014rva,Lourenco:2020qft}  in which $r_{\rm skin}^{208}$ and/or $L$ is presented. 
The correlation between $r_{\rm skin}^{208}$ and $L$ is more reliable when the number of EoSs is larger. 
The resulting relation  
\bea
L=620.39~r_{\rm skin}^{208}-57.963~{\rm MeV}
\label{Eq:skin-L}
\eea
has a strong correlation with the  correlation coefficient $R=0.99$, as shown in Fig.~\ref{Fig-L-skin}.

The relation~\eqref{Eq:skin-L} allows us to deduce a constraint on $L$ from 
the PREX2 value. 
The resulting  range of $L$ are $L = 76 \text{--} 165$ MeV, while equation shown in Ref.~\cite{RocaMaza:2011pm} yields 
$L = 76\text{--} 172$ MeV.
These values and $L =69 \text{--} 143 ~{\rm MeV}$  support stiffer EoSs. 
The stiffer EoSs allow us to consider the phase transition such as QCD transition in NS. 
The following EoSs satisfy $L = 76 \text{--} 172$ MeV;
SkO,
FKVW,
Rs,
SV-sym34,
es325,
TFa,
NL$\rho$,
BSR6,
R$\sigma$,
Sk-Rs,
E0009,
Gs,
Z271,
GM3,
PKDD,
E0008(TMA),
SK272,
GM1,
G$\sigma$,
Sk-T4,
SK255,
SV,
es35,
S271,
SkI3,
rG2,
PC-PK1,
SkI2,
E0025,
PC-LA,
rNLC,
E0036,
rTM1,
TM1,
NL4,
rNL-SH,
NL-SH,
rNL-RA1,
PC-F2,
PK1,
PC-F1,
NL3,
rNL3,
PC-F3,
PC-F4,
NL3*,
TFb,
rNL3*,
SkI5,
NL2,
rNL-Z,
TFc,
NL1,
rNL1,
SkI1 in Table I of  Ref.~\cite{TAGAMI2022105155}.

As an indirect measurement, meanwhile, the high-resolution $E1$ polarizability experiment ($E1$pE) yields 
\bea
r_{\rm skin}^{208}(E1{\rm pE}) &=&0.156^{+0.025}_{-0.021}=0.135\text{--}0.181~{\rm fm} 
\label{Eq:skin-Pb208-E1}
\eea
for $^{208}$Pb~\cite{Tamii:2011pv} 
\bea
r_{\rm skin}^{48}(E1{\rm pE}) &=&0.17 \pm 0.03=0.14\text{--}0.20~{\rm fm}~~~  
\label{Eq:skin-Ca48-E1}
\eea
for $^{48}$Ca~\cite{Birkhan:2016qkr}. 

There is no overlap between $r_{\rm skin}^{208}({\rm PREX2})$ and  $r_{\rm skin}^{208}(E1{\rm pE})$ 
in one  $\sigma$ level. 
However, we determined a value of $r_{\rm skin}^{208}({\rm exp})$ 
from measured reaction cross sections $\sigma_{\rm R}({\rm exp})$ of 
p+$^{208}$Pb scattering in a range of incident energies, $30 \lsim E_{\rm in} \lsim 100$~MeV~\cite{Tagami:2020bee}; 
the value is $r_{\rm skin}^{208}({\rm exp})=0.278 \pm 0.035$~fm. Our result agrees 
with $R_{\rm skin}^{208}({\rm PREX2})$. We also deduced    
$r_{n}({\rm exp})=5.722 \pm 0.035$~fm and $r_{\rm m}({\rm exp})=5.614 \pm 0.022$~fm in addition to 
$r_{\rm skin}^{208}({\rm exp})$. 
As for He+$^{208}$Pb scattering, we determine $r_{\rm skin}^{208}({\rm exp})=0.416 \pm 0.146$~fm 
\cite{Matsuzaki:2021hdm}.
Our results are consistent with PREX II and therefore supports larger slope parameter $L$. 

Our model is the chiral (Kyushu) $g$-matrix folding model with the densities calculated
with Gogny-D1S HFB (D1S-GHFB) 
with the angular momentum projection (AMP)~\cite{Toyokawa:2017pdd,PRC.101.014620}. 
For p+$^{208}$Pb scattering, the neutron  density is scaled so that the $r_{n}$ 
of the scaled neutron density 
can reproduce the data~\cite{Carlson:1975zz,Ingemarsson:1999sra,Auce:2005ks} on $\sigma_{\rm R}$, 
since the $r_{\rm p}$ of D1S-GHFB+AMP proton density agrees 
with  the $r_p({\rm exp})$~\cite{PRC.90.067304} determined from electron scattering. 
For $^{12}$C scattering on $^{9}$Be, $^{12}$C, $^{27}$Al 
targets, we tested reliability of the Kyushu $g$-matrix folding model  and found that 
the Kyushu $g$-matrix folding model is reliable in $30 \lsim E_{\rm in} \lsim 100$~MeV and 
$250 \lsim E_{\rm in} \lsim 400$~MeV~\cite{PRC.101.014620}. 
This is the reason why we took  $30 \lsim E_{\rm in} \lsim 100$~MeV in the analyses~\cite{Tagami:2020bee} 
of p+$^{208}$Pb scattering. 
After the analyses, we find  that the Kyushu $g$-matrix folding model reproduces the lower bound of   
the data on $\sigma_{\rm R}$~\cite{Kox:1985ex} for $^{12}$C+$^{12}$C scattering at $E_{\rm in}=10.4$~MeV per nucleon.

The $g$-matrix folding model is a standard way of deriving the microscopic optical potential 
for proton scattering and nucleus-nucleus scattering~\cite{NPA.291.299,*NPA.291.317,*NPA.297.206,PR.55.183,*Satchler83,PTP.70.459,*PTP.73.512,*PTP.76.1289,ANP.25.275,%
PRC.78.044610,*PRC.79.011601,*PRC.80.044614,PRC.89.064611,JPG.42.025104,*JPG.44.079502,PRC.92.024618,*PRC.96.059905,%
PTEP.2018.023D03,PRC.101.014620,PRL.108.052503}.  
The folding model is composed of the single folding model for  proton scattering and the double folding model 
for  nucleus-nucleus scattering. 
The relation between the single and the double folding model is clearly shown in Ref.~\cite{PRC.89.064611}.
Applying the  double-folding model based on Melbourne $g$-matrix~\cite{ANP.25.275} 
for the data~\cite{Takechi:2012zz} on interaction cross sections,     
we found that $^{31}$Ne is a halo nucleus with large deformation~\cite{PRL.108.052503},   
and deduced the matter radii $r_{\rm m}$ for Ne isotopes~\cite{PRC.85.064613}. 
Also for Mg isotopes, we determined the  $r_{\rm m}$ from $\sigma_{\rm R}({\rm exp})$ 
for scattering of Mg isotopes on a $^{12}$C target~\cite{PRC.89.044610}.

Now, we consider proton scattering on $^{208}$Pb, $^{58}$Ni, $^{40,48}$Ca, $^{12}$C targets, 
since there is no interaction cross section for proton scattering. 
In fact, good data on $\sigma_{\rm R}$ 
are available in Refs.~\cite{Carlson:1975zz,Ingemarsson:1999sra,Auce:2005ks} for $^{208}$Pb, 
Refs~\cite{Auce:2005ks,Ingemarsson:1999sra,EliyakutRoshko:1995fn,Dicello:1967zz,Bulman:1968ujl} for 
$^{58}$Ni, Ref.~\cite{Carlson:1994fq} for $^{48}$Ca, 
Refs.~\cite{Carlson:1975zz,Ingemarsson:1999sra,Auce:2005ks} for $^{40}$Ca, and 
Refs.~\cite{Auce:2005ks,Ingemarsson:1999sra,Menet:1971zz} for $^{12}$C.
We have already shown that 
for p+$^{208}$Pb scattering the $\sigma_R$ calculated with $r_{\rm skin}^{208}({\rm PREX2})$ and 
$r_{\rm p}({\rm exp})=5.444$~fm~\cite{PRC.90.067304} of electron scattering
reproduce the data at $E_{\rm lab} = 534.1, 549, 806$ MeV~\cite{WAKASA2021104749}.

In this paper, we first test the Kyushu $g$-matrix single folding model for p+$^{208}$Pb scattering, 
since the PREX2 data is available.  
We find that the present model is reliable in  $20 \lsim E_{\rm in} \lsim 180$~MeV, as shown 
in Sec.~\ref{$^{208}$Pb}. 
After the testing, we determine 
 $r_{\rm m}({\rm exp})$, $r_{\rm n}({\rm exp})$,  $r_{\rm skin}({\rm exp})$ for $^{208}$Pb, 
 $^{58}$Ni, $^{40,48}$Ca, $^{12}$C
from the $\sigma_{\rm R}({\rm exp})$ in  $20 \lsim E_{\rm in} \lsim 180$~MeV, 
as shown in Sec.~\ref{$^{58}$Ni,$^{48,40}$Ca,$^{12}$C}. 
For each nucleus, the D1S-GHFB+AMP proton and neutron densities are scaled so as to reproduce 
$\sigma_{\rm R}({\rm exp})$ under that condition that the $r_{\rm p}({\rm scaling})$ of the scaled proton density agrees with $r_{\rm p}({\rm exp})$ of electron scattering. 

We explain our model in Sec.~\ref{Sec-Framework} and our results in Sec.~\ref{Results}. 
Section \ref{Summary} is devoted to a summary. 

\section{Model}
\label{Sec-Framework}

Our model is the Kyushu $g$-matrix  folding model~\cite{Toyokawa:2017pdd,PRC.101.014620} 
with the proton and neutron densities scaled from the D1S-GHFB+AMP densities.  

\subsection{The Kyushu $g$-matrix folding model}
\label{The Kyushu $g$-matrix folding model}

\begin{figure}[H]
\begin{center}
 \includegraphics[width=0.45\textwidth,clip]{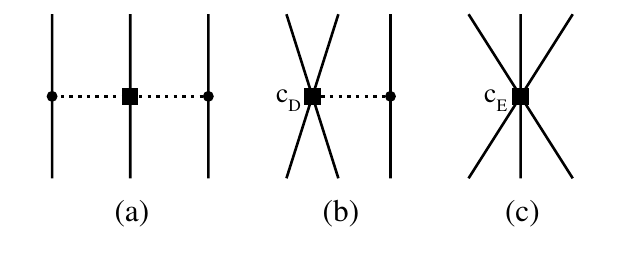}
 \caption{3$N$Fs in NNLO (next-to-next-to-leading-order). 
Diagram (a) corresponds 
to the Fujita-Miyazawa 2$\pi$-exchange 3$N$F (next-to-next-to-next-to-leading-order) 
\cite{PTP.17.360,*PTP.17.366}, 
and diagrams (b) and (c) correspond to 1$\pi$-exchange and contact 3$N$Fs.
 The solid and dashed lines denote nucleon and pion propagations, 
respectively, and filled circles and squares stand for vertices. 
 The strength of the filled-square vertex is often called $c_{D}$ 
in diagram (b) and $c_{E}$ in diagram (c). 
}
 \label{fig:diagram}
\end{center}
\end{figure}

Kohno calculated the $g$ matrix  for the symmetric nuclear matter, 
using the Brueckner-Hartree-Fock method with chiral N$^{3}$LO 2NFs and NNLO 3NFs~\cite{PRC.88.064005,*PRC.96.059903}, where N$^{3}$LO 3NF is abbreviation of 
next-to-next-to-next-to-leading-order three-body force and 
NNLO  2NFs is of next-to-next-to-leading-order two-body force. 
He set $c_D=-2.5$ and $c_E=0.25$ so that  the energy per nucleon can  become minimum 
at $\rho = \rho_{0}$; see Fig.~\ref{fig:diagram} for the definition of $c_D$ and $c_E$.
Toyokawa {\it et al.} localized the non-local chiral  $g$ matrix into three-range Gaussian forms~\cite{
Toyokawa:2017pdd}, using the localization method proposed 
by the Melbourne group~\cite{ANP.25.275,PRC.44.73,PRC.49.1309}. 
The resulting local  $g$ matrix is referred to as  ``Kyushu  $g$-matrix''. 
The Kyushu $g$-matrix is constructed from chiral interaction with the cutoff 550~MeV.  
fig-chiral.epsfig-chiral

The  Kyushu $g$-matrix folding model is successful in reproducing $\sigma_{\rm R}$, 
differential cross sections  $d\sigma/d\Omega$, vector analyzing powers $A_y$ for $^4$He scattering 
 in $E_{\rm in}=30 \text{--} 200$~MeV per nucleon~\cite{Toyokawa:2017pdd}. 
The success is true for proton scattering at $E_{\rm in}=65$~MeV~\cite{Toyokawa:2014yma}.

In Ref.~\cite{Tagami:2019svt}, we tested the Kyushu $g$-matrix folding model~\cite{Toyokawa:2017pdd} 
for  $^{12}$C scattering on  $^{9}$Be, $^{12}$C, $^{27}$Al targets 
 in  $30  \lsim E_{\rm in} \lsim 400 $~MeV. We found that 
the Kyushu $g$-matrix folding model is reliable for $\sigma_{\rm R}$ 
in $30  \lsim E_{\rm in} \lsim 100 $~MeV and $250  \lsim E_{\rm in} \lsim 400$~MeV.  
This indicates that the Kyushu $g$-matrix folding model is applicable in the $E_{\rm in}$ range.  
After the test, we found that our model reproduces the lower bound of measured reaction cross 
section $\sigma_{\rm R}$ \cite{Kox:1985ex} at $E_{\rm in}=10.4$~MeV. 
Our model is reliable for $10  \lsim E_{\rm in} \lsim 100 $~MeV and $250  \lsim E_{\rm in} \lsim 400$~MeV.

We recapitulate the single folding model for nucleon-nucleus scattering. 
The potential $U(\vR)$ consists of the direct and exchange parts~\cite{PRC.89.064611},
$U^{\rm DR}(\vR)$ and $U^{\rm EX}(\vR)$, defined by 
\begin{subequations}
\begin{eqnarray}
U^{\rm DR}(\vR) & = & 
\sum_{\mu,\nu}\int             \rho^{\nu}_{\rm T}(\vrr_{\rm T})
            g^{\rm DR}_{\mu\nu}(s;\rho_{\mu\nu})  d
	    \vrr_{\rm T}\ ,\label{eq:UD} \\
U^{\rm EX}(\vR) & = & 
\sum_{\mu,\nu}
\int \rho^{\nu}_{\rm T}(\vrr_{\rm T},\vrr_{\rm T}+\vs) \nonumber \\
                &   &
\times g^{\rm EX}_{\mu\nu}(s;\rho_{\mu\nu}) \exp{[-i\vK(\vR) \cdot \vs/M]}
             d \vrr_{\rm T}\ ,\label{eq:UEX}
\end{eqnarray}
\end{subequations}
where $\vR$ is the relative coordinate between a projectile (P)  and 
a target (${\rm T}$),
$\vs=-\vrr_{\rm T}+\vR$, and $\vrr_{\rm T}$ is
the coordinate of the interacting nucleon from the center-of-mass of T.
Each of $\mu$ and $\nu$ denotes the $z$-component of isospin, i.e., 
$(1/2,-1/2)$ corresponds to (neutron, proton).
 The nonlocal $U^{\rm EX}$ has been localized in Eq.~\eqref{eq:UEX}
with the local semi-classical approximation
\cite{NPA.291.299,*NPA.291.317,*NPA.297.206},
where \vK(\vR) is the local momentum between P and T, 
and $M= A/(1 +A)$ for the target mass number $A$;
see Ref.~\cite{Minomo:2009ds} for the validity of the localization.
 The direct and exchange parts, $g^{\rm DR}_{\mu\nu}$ and
$g^{\rm EX}_{\mu\nu}$, of the $g$-matrix depend on the local density
\bea
 \rho_{\mu\nu}=\sigma^{\mu} \rho^{\nu}_{\rm T}(\vrr_{\rm T}+\vs/2)
\label{local-density approximation}
\eea
at the midpoint of the interacting nucleon pair, where $\sigma^{\mu}$ having ${\mu}=-1/2$
is the Pauli matrix of an 
incident proton. As a way of taking the center-of-mass correction to 
the D1S-GHFB+AMP densities,  we use the method of Ref.~\cite{PRC.85.064613}, 
since the procedure is quite simple. 

The direct and exchange parts, $g^{\rm DR}_{\mu\nu}$ and 
$g^{\rm EX}_{\mu\nu}$, of the $g$-matrix,  are described by~\cite{PRC.85.064613}
\begin{align}
&\hspace*{0.5cm} g_{\mu\nu}^{\rm DR}(s;\rho_{\mu\nu}) \nonumber \\ 
&=
\begin{cases}
\displaystyle{\frac{1}{4} \sum_S} \hat{S}^2 g_{\mu\nu}^{S1}
 (s;\rho_{\mu\nu}) \hspace*{0.42cm} ; \hspace*{0.2cm} 
 {\rm for} \hspace*{0.1cm} \mu+\nu = \pm 1 
 \vspace*{0.2cm}\\
\displaystyle{\frac{1}{8} \sum_{S,T}} 
\hat{S}^2 g_{\mu\nu}^{ST}(s;\rho_{\mu\nu}), 
\hspace*{0.2cm} ; \hspace*{0.2cm} 
{\rm for} \hspace*{0.1cm} \mu+\nu = 0 
\end{cases}
\\
&\hspace*{0.5cm}
g_{\mu\nu}^{\rm EX}(s;\rho_{\mu\nu}) \nonumber \\
&=
\begin{cases}
\displaystyle{\frac{1}{4} \sum_S} (-1)^{S+1} 
\hat{S}^2 g_{\mu\nu}^{S1} (s;\rho_{\mu\nu}) 
\hspace*{0.34cm} ; \hspace*{0.2cm} 
{\rm for} \hspace*{0.1cm} \mu+\nu = \pm 1 \vspace*{0.2cm}\\
\displaystyle{\frac{1}{8} \sum_{S,T}} (-1)^{S+T} 
\hat{S}^2 g_{\mu\nu}^{ST}(s;\rho_{\mu\nu}) 
\hspace*{0.2cm} ; \hspace*{0.2cm}
{\rm for} \hspace*{0.1cm} \mu+\nu = 0 ~~~~~
\end{cases}
\end{align}
where $\hat{S} = {\sqrt {2S+1}}$ and $g_{\mu\nu}^{ST}$ are 
the spin-isospin components of the $g$-matrix; see Ref.~\cite{PTEP.2018.023D03} for the explicit form of 
$g^{\rm DR}_{\mu\nu}$ and $g^{\rm EX}_{\mu\nu}$.

The potential   $U(\vR)$ thus obtained has the form of 
$U(\vR)=U_{\rm cent}(R)+{\hat {\ell}} \cdot {\hat \sigma}~U_{\rm spin-orbit}(R)$, where 
$U_{\rm cent}(R)$ and   $U_{\rm spin-orbit}(R)$ are the central and the spin-orbit part of $U(\vR)$, respectively, 
and  $\ell$ is the orbital angular momentum of the proton scattering; see Eq. (28) 
in Ref.~\cite{Sumi:2012fr} for the derivation. 
The relative wave function $\psi$ between P and T can be decomposed into partial waves $\chi_{\ell}$,
each with different  $\ell$. The $\chi_{\ell}$ is obtained by solving the Schr\"{o}dinger 
equation having  $U(\vR)$. 
The elastic $S$-matrix elements $S_{\ell}$ are obtained from the asymptotic form of
$\chi_{\ell}$. The total reaction cross section $\sigma_{\rm R}$ is calculable from
the $S_{\ell}$ as
\bea
\sigma_{\rm R}=\frac{\pi}{K^2}\sum_{\ell} (2{\ell}+1)(1-|S_{\ell}|^2).
\eea

\subsection{Scaling procedure of proton and neutron densities}

The proton and neutron densities, $\rho_p(r)$ and $\rho_n(r)$, 
are scaled from the D1S-GHFB+AMP densities. 
We can obtain the scaled density $\rho_{\rm scaling}(\vrr)$ from the original density $\rho(\vrr)$ as
\bea
\rho_{\rm scaling}(\vrr)=\frac{1}{\a^3}\rho(\vrr/\a)
\eea
with a scaling factor
\bea
\a=\sqrt{ \frac{\langle \vrr^2 \rangle_{\rm scaling}}{\langle \vrr^2 \rangle}} .
\eea
For later convenience, we refer to the proton (neutron) radius of the scaled proton (neutron) density $\rho^{\rm p}_{\rm scaling}(\vrr)$ 
($\rho^{\rm n}_{\rm scaling}(\vrr)$) as $r_{\rm p}({\rm scaling})$ ($r_{\rm n}({\rm scaling})$).

Table \ref{reference values-a} shows the scaling factors, $\a_{\rm p}$ and $\a_{\rm n}$,  
from the D1S-GHFB+AMP densities to 
the scaled densities that reproduce data $\sigma_{\rm R}({\rm exp})$ and $r_{\rm p}({\rm exp})$ of electron scattering.

\begin{table}[htb]
\begin{center}
\caption
{Scaling factors $\a_{\rm n}$ and $\a_{\rm p}$ for neutron and proton.
 }
\begin{tabular}{ccc}
\hline\hline
 & $\a_{\rm n}$ & $\a_{\rm p}$\\
\hline
 $^{208}$Pb & $1.015 $ & $1.000 $ \\
 $^{58}$Ni  & $1.003 $ & $0.994  $ \\
 $^{48}$Ca & $0.973 $ & $0.982 $ \\
 $^{40}$Ca & $1.000$ & $0.995$  \\
 $^{12}$C  & $0.942 $ & $0.957$ \\
\hline
\end{tabular}
 \label{reference values-a}
 \end{center} 
 \end{table}

\subsection{Effective nucleon-nucleon interaction for targets}

For a change of the proton and neutron distributions in targets, 
a microscopic approach is to modify  D1S. 
For the Gogny EoSs, the effective nucleon-nucleon interaction can be described as   
\bea
~~~~~~V(\vec{r})  &=&\sum_{i=1,2} t_0^i (1+x_0^i P_\sigma) \rho^{\alpha_i} \delta (\vec{r})
  \nonumber \\
 &+&\sum_{i=1,2} (W_i+B_i P_\sigma -H_i P_\tau -M_i P_\sigma P_\tau) e^{-\frac{r^2}{\mu_i^2}}
  \nonumber \\
 &+&i W_0 (\sigma_1 +\sigma_2) [\vec{k'} \times \delta (\vec{r}) \vec{k}], 
\eea 
where $\sigma$ and $\tau$ are the Pauli spin and isospin operators,
respectively, and the corresponding exchange operators $P_{\sigma}$ and $P_{\tau}$ are defined as usual.

For  $^{208}$Pb, we have changed all parameters of D1S, but cannot find the NN interaction that reproduce 
$r_{\rm skin}^{208}({\rm PREX2}) = 0.283\pm 0.071\,{\rm fm}$. The best fitting is  
the D1PK2-GHFB+AMP with $r_{\rm skin}^{208}({\rm D1PK2}) = 0.185\,{\rm fm}$; 
note that $r_{\rm skin}^{208}({\rm D1S}) = 0.137\,{\rm fm}$. 
The parameters of D1PK2 are shown in Table \ref{paramters:D1PK2}.

\squeezetable
\begin{table}[h]
\caption{
Parameter sets of D1PK2. 
}
 \begin{tabular}{c|c|cccc|ccc|c}
\hline
  D1PK &$\mu_i$&$W_i$ & $B_i$ & $H_i$ & $M_i$ &
  $t_0^i$ & $x_0^i$ & $\alpha_i$& $W_0$\cr
\hline
   $i=1$ & 0.90 &  -465.027582 & 155.134492 & -506.775323& 117.749903&
  981.065351& 1 & 1/3  & 130\cr
  $i=2$ & 1.44 &34.6200000& -14.0800000& 70.9500000& -41.3518104&
  534.155654 & -1 & 1 & \cr
\hline
\end{tabular}
\label{paramters:D1PK2}
\end{table}  


\section{Results}
\label{Results}

First of all, we regard reliable $r_{\rm m}$ and $r_{\rm n}$ as reference values, 
$r_{\rm m}({\rm ref})$ and $r_{\rm n}({\rm ref})$, 
in order to determine $r_{\rm m}({\rm exp})$ from  $\sigma_{\rm R}({\rm exp})$.
The reference values are shown below; see 
Table  \ref{reference values} for the numerical values. 
Whenever we calculate $\sigma_{\rm R}$, we use the Kyushu $g$-matrix model.

\begin{description}

\item[ $^{58}$Ni]
The reference values are the $r_{\rm m}({\rm AMP})$ and $r_{\rm n}({\rm AMP})$ calculated with D1S-GHFB+AMP,  
since the $\sigma_{\rm R}({\rm AMP})$  
are near the upper bound of the data, 
as shown in Fig.~\ref{Fig-RXsec-p+Ni58}. 
$E_{\rm in}$ dependence of $\sigma_{\rm R}({\rm AMP})$ 
is similar to that of the data. 
We then define the ratio $F(E_{\rm in}) \equiv \sigma_{\rm R}({\rm exp})/\sigma_{\rm R}({\rm ref})=
\sigma_{\rm R}({\rm exp})/\sigma_{\rm R}({\rm AMP})$, and  introduce  the average value of $F(E_{\rm in})$ as a fine-tuning factor $F$. 
The factor is $F=0.96473$ close to 1. 
The $F \sigma_{\rm R}({\rm AMP})$ almost reproduce the central values of the  data, 
as shown in Fig.~\ref{Fig-RXsec-p+Ni58}.  
The scaling procedure is not made to get the reference values, i.e., $\a=1$. 

\item[ $^{208}$Pb]
The reference values are the $r_{\rm m}({\rm PREX2})$ and $r_{\rm n}({\rm PREX2})$ evaluated  from $r_{\rm skin}^{208}({\rm PREX2})$ 
and  $r_{\rm p}({\rm exp})$~\cite{PRC.90.067304} of electron scattering. 
In this case, the $\sigma_{\rm R}({\rm exp})$ based on the densities  scaled to
 $r_{\rm n}({\rm PREX2})$ and  $r_{\rm p}({\rm exp})$ 
reproduce the data within error-bar, 
as shown in Fig.~\ref{Fig-RXsec-p+Pb208-2}. 
For this reason, $F$ is 1.

\item[ $^{48}$Ca]
We can obtain $r_{\rm m}({\rm CREX})$  and $r_{\rm n}({\rm CREX})$ 
from the CREX value of Eq.~\eqref{CREX-value} and $r_{\rm p}({\rm exp})=3.385~{\rm fm}$~\cite{Angeli:2013epw} of electron scattering.
In this case, the $\sigma_{\rm R}({\rm CREX})$ based on $r_{\rm n}({\rm CREX})$ and  $r_{\rm p}({\rm exp})$ 
are near the central values of of the data, 
as shown in Fig.~\ref{Fig-RXsec-p+Ca48}. 
The fine-tuning factor is $F=0.9810$ close to 1. 
The $F \sigma_{\rm R}(E1{\rm pE})$ almost reproduce the central values of the  data, 
as shown in Fig.~\ref{Fig-RXsec-p+Ca48}. 

\item[ $^{40}$Ca]
As for $^{40,48}$Ca, Zenihiro {\it et al.} measured 
the differential cross section and the analyzing powers for p+$^{40,48}$Ca scattering in RCNP, 
and determined $r_{\rm skin}^{40,48}({\rm RCNP})$~\cite{Zenihiro:2018rmz}.  
For $^{48}$Ca, the value $r_{\rm skin}({\rm RCNP})=0.168^{+0.025}_{-0.028}$~fm 
is consistent with $r_{\rm skin}^{48}(E1{\rm pE})$.  
For $^{40}$Ca, their values are shown in Table  \ref{reference values} as reference values. 
Since $\sigma_{\rm R}({\rm RCNP})=\sigma_{\rm R}({\rm AMP})$, the  $\sigma_{\rm R}({\rm AMP})$ are reliable.
In this case, the $\sigma_{\rm R}({\rm AMP})$ overshoot the data,  as shown in Fig.~\ref{Fig-RXsec-p+Ca40}. 
The fine-tuning factor is $F=0.92716$ close to 1. 
The $F \sigma_{\rm R}({\rm AMP})$ almost reproduce the central values of the  data, 
as shown in Fig.~\ref{Fig-RXsec-p+Ca40}. 

\item[ $^{12}$C]
Tanihata {\it et. al.} measured interaction cross sections at 790 MeV/nucleon in GSI and determined 
$r_{\rm m}({\rm GSI})=2.35$~fm for $^{12}$C~\cite{Tanihata:1988ub}. 
The $r_{\rm m}({\rm GSI})$ and $r_{\rm p}({\exp})=2.327$~fm~\cite{Angeli:2013epw}
lead to $r_{\rm n}({\rm GSI})=2.37$~fm.
The $r_{\rm m}({\rm GSI})$ and $r_{\rm n}({\rm GSI})$ are  reference values. 
The fine-tuning factor is $F=0.93077$ close to 1. 
The $F \sigma_{\rm R}({\rm GSI})$ are near the central values of 
$\sigma_{\rm R}({\rm exp})$, as shown in Fig.~\ref{Fig-RXsec-p+C12}.

\end{description}

\squeezetable
\begin{table}[htb]
\begin{center}
\caption
{Reference values of   $r_{\rm m}({\rm ref})$,  $r_{\rm n}({\rm ref})$, $r_{\rm skin}({\rm ref})$ 
together with $r_{\rm p}({\rm exp})$ of electron scattering. 
The $r_{\rm p}({\rm exp})$ are deduced from the electron scattering~\cite{Angeli:2013epw,PRC.90.067304}. 
In actual calculations, the central values are taken as reference values. 
The radii are shown in units of fm.  
 }
\begin{tabular}{cccccc}
\hline\hline
 & Ref. & $r_{\rm p}({\rm exp})$ & $r_{\rm m}({\rm ref})$ &  $r_{\rm n}({\rm ref})$ & $r_{\rm skin}({\rm ref})$ \\
\hline
 $^{208}$Pb & PREX2 & $5.444$ & $5.617 \pm 0.044$ & $5.727 \pm 0.071$ & $0.283\pm 0.071$ \\
 $^{58}$Ni & D1S AMP & $3.727$ & $3.721 $ & $3.715 $ & $-0.013 $ \\
 $^{48}$Ca & CREX & $3.385$ & $3.456 \pm 0.050$ & $3.506 \pm 0.050$ & $0.121 \pm 0.050$ \\
 $^{40}$Ca & \cite{Zenihiro:2018rmz} & $3.385$ & $3.380^{+0.022}_{-0.023}$ & $3.375^{+0.022}_{-0.023}$ & $-0.010^{+0.022}_{-0.023}$ \\
 $^{12}$C & \cite{Tanihata:1988ub} & $2.327$ & $2.35 \pm 0.02 $ & $2.37  \pm 0.02$ & $0.05  \pm 0.02$ \\
\hline
\end{tabular}
 \label{reference values}
 \end{center} 
 \end{table}

\begin{table}[htb]
\begin{center}
\caption
{Fine-tuning factor $F$. 
 }
\begin{tabular}{cc}
\hline\hline
  & $F$ \\
\hline
 $^{208}$Pb  & $1 $\\
 $^{58}$Ni  & $0.96473 $\\
 $^{48}$Ca  & $0.9810 $\\
 $^{40}$Ca  & $0.92716 $ \\
 $^{12}$C  & $0.93077$\\
\hline
\end{tabular}
 \label{reference values-a}
 \end{center} 
 \end{table}

When we determine  $r_{\rm m}({\rm exp})$ from data $\sigma_{\rm R}({\rm exp})$, 
we scale the D1S-GHFB+AMP 
proton and neutron densities so as to $F \sigma_{\rm R}({\rm ref})=\sigma_{\rm R}({\rm exp})$ and  
$r_{\rm p}({\rm scaling})=r_{\rm p}({\rm exp})$. 
Next, we deduce $r_{\rm m}({\rm exp})$ from
 $r_{\rm n}({\rm scaling})$ and $r_{\rm p}({\rm scaling})$ for each $E_{\rm in}$.  
 The resulting $r_{\rm m}({\rm exp})$ depends on $E_{\rm in}$.
For all the $r_{\rm m}({\rm exp})$, we take the weighted mean and its error. 
Finally, we evaluate $r_{\rm skin}({\rm exp})$ and $r_{\rm n}({\rm exp})$ from the resulting 
$r_{\rm m}({\rm exp})$ and the $r_{\rm p}({\rm exp})$~\cite{Angeli:2013epw,PRC.90.067304} of the electron scattering. 
For later convenience, we refer to this procedure as ``experimental scaling procedure with $F$ (ESP-F)''.

\subsection{Test of the Kyushu $g$-matrix folding model for p+$^{208}$Pb scattering}
\label{$^{208}$Pb}

\begin{figure}[H]
\begin{center}
 \includegraphics[width=0.45\textwidth,clip]{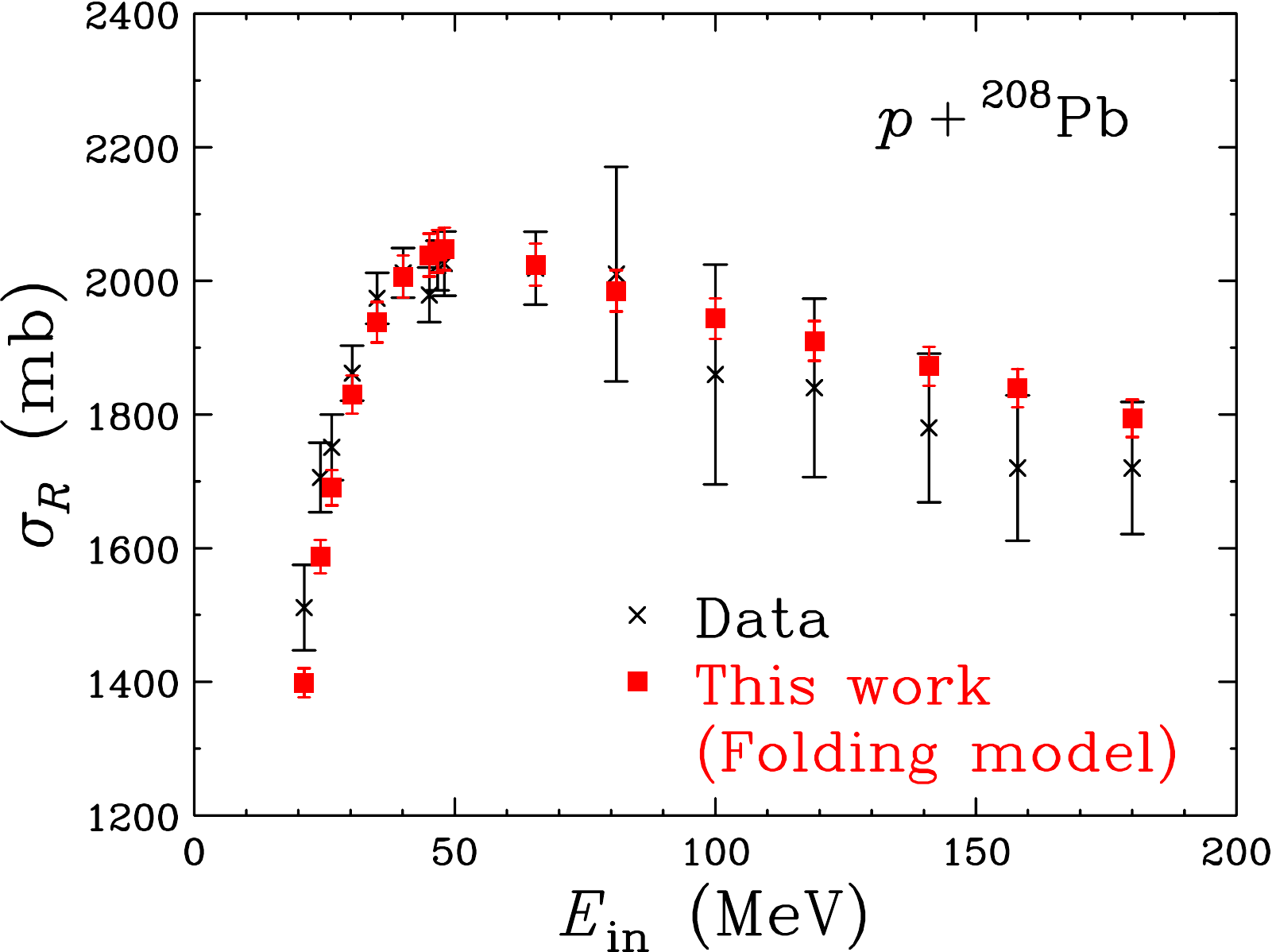}
  \caption{ 
 $E_{\rm in}$ dependence of reaction cross sections $\sigma_{\rm R}$ 
 for $p$+$^{208}$Pb scattering. 
The squares with error bar (legend ``This work (Folding model)'') stand for the results 
of the folding model with the densities scaled to PREX2, 
whereas the crosses with error bar correspond to 
the data~\cite{Carlson:1975zz,Ingemarsson:1999sra,Auce:2005ks} on $\sigma_{\rm R}$ 
    }
 \label{Fig-RXsec-p+Pb208-2}
\end{center}
\end{figure}

Now we test the Kyushu $g$-matrix model for proton scattering. 
As shown in  Fig.~\ref{Fig-RXsec-p+Pb208-2}, 
the $\sigma_{\rm R}$ (squares with error bar)  based on 
$R_{\rm skin}^{208}({\rm PREX2})$ and $r_{\rm p}({\rm exp})$
are consistent with data~\cite{Carlson:1975zz,Ingemarsson:1999sra,Auce:2005ks} 
in $20  \lsim E_{\rm in} \lsim 180$ MeV; see Table  \ref{reference values} for 
$R_{\rm skin}^{208}({\rm PREX2})$ and $r_{\rm p}({\rm exp})$.
This indicates that our model is good in $20  \lsim E_{\rm in} \lsim 180$ MeV for proton scattering.

\subsection{$^{208}$Pb,$^{58}$Ni,$^{48,40}$Ca,$^{12}$C}
\label{$^{58}$Ni,$^{48,40}$Ca,$^{12}$C}

\subsubsection{$^{208}$Pb}

\begin{figure}[H]
\begin{center}
  \includegraphics[width=0.45\textwidth,clip]{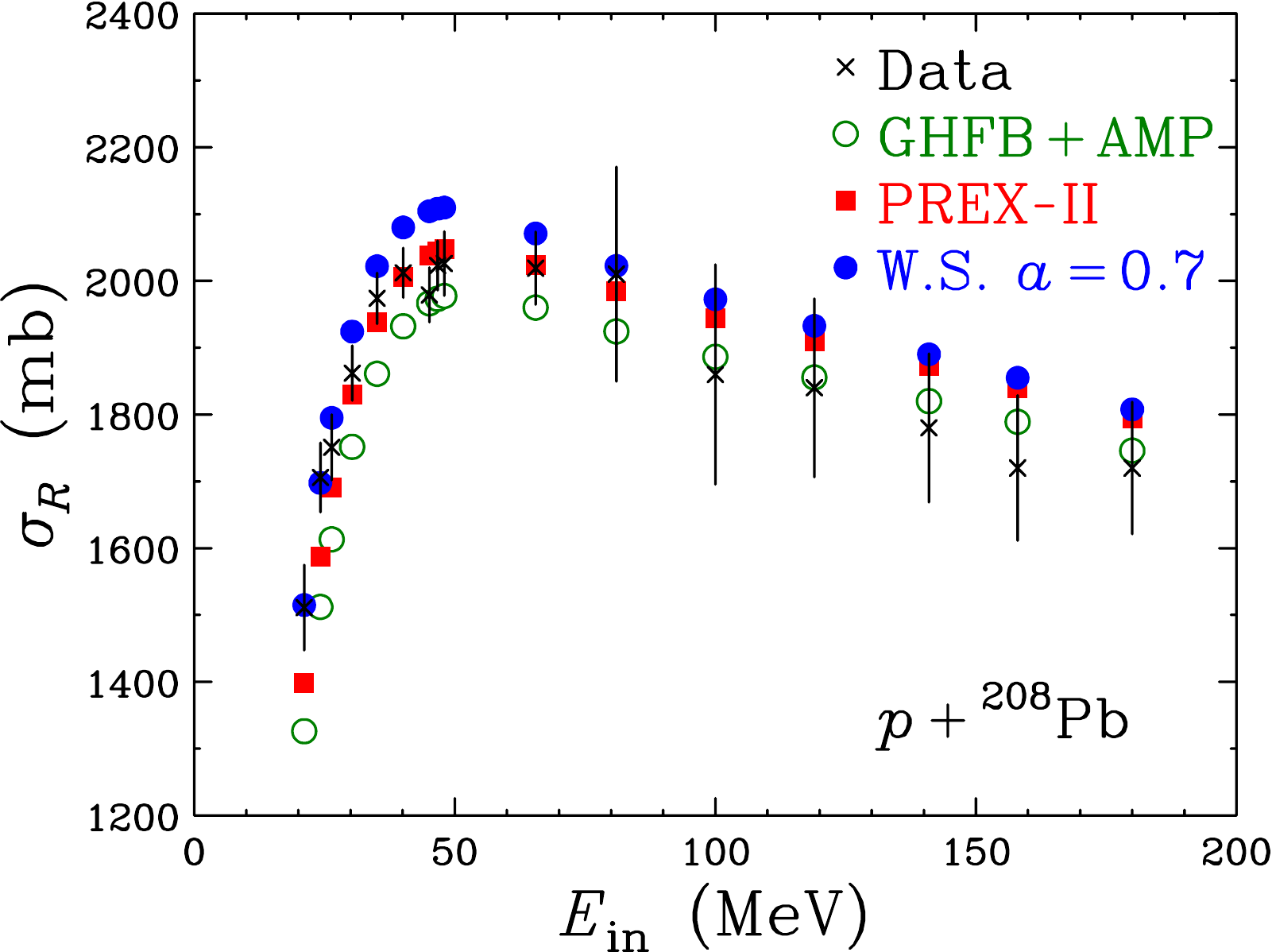}
  \caption{ 
 $E_{\rm in}$ dependence of reaction cross sections $\sigma_{\rm R}$ 
 for $p$+$^{208}$Pb scattering. 
Squares stand for the results  
of the folding model with the neutron  density scaled to the central value of PREX2. 
Open circles denote the results of the D1S-GHFB+AMP densities, and close circles 
correspond to the results of the Woods-Saxon type neutron density ($r_{\ WS}=6.59$~fm, $a_{\ WS}=0.7$~fm) 
fitted to the central value of PREX2. 
As for the proton density, it is calculated with D1S-GHFB+AMP for three types of calculations. 
The crosses with error bar are   
the data~\cite{Carlson:1975zz,Ingemarsson:1999sra,Auce:2005ks} on 
$\sigma_{\rm R}$ 
 }
 \label{Fig-RXsec-p+Pb208-3}
\end{center}
\end{figure}

Figure~\ref{Fig-RXsec-p+Pb208-3} shows reaction cross sections 
 $\sigma_{\rm R}$ of $p$+$^{208}$Pb scattering as a 
function of  $E_{\rm in}$. The results of the D1S-GHFB+AMP densities reproduce 
the data~\cite{Carlson:1975zz,Ingemarsson:1999sra,Auce:2005ks} with 4\% errors. 
This is true for the neutron density scaled to the central value of PREX2 and 
the D1S-GHFB+AMP density. 
In the results of 
the Woods-Saxon type neutron density ($r_{\ WS}=6.59$~fm, $a_{\ WS}=0.7$~fm) 
fitted to the central value of PREX2, 
we use the D1S-GHFB+AMP proton density,  
The results of the Woods-Saxon type neutron density($r_{\ WS}=6.59$~fm, $a_{\ WS}=0.7$~fm) and 
the D1S-GHFB+AMP proton density 
are close to those of the neutron density scaled to the central value of PREX2 and 
the D1S-GHFB+AMP proton density. 
The Woods-Saxon type neutron density ($r_{\ WS}=6.81$~fm, $a_{\ WS}=0.6$~fm)
yields almost the same results as the case of $r_{\ WS}=6.59$~fm, $a_{\ WS}=0.7$~fm, that is, 
the former  undershoots the latter by 0.974. 
We then do not show the former results in Fig.~\ref{Fig-RXsec-p+Pb208-3}.

The results of ESP-F are $r_{\rm skin}^{208}({\rm exp})=0.299 \pm 0.020$~fm, 
$r_{\rm n}({\rm exp})=5.743 \pm 0.020$~fm, $r_{\rm m}({\rm exp})=5.627 \pm 0.020$~fm. 
The present skin value $0.299 \pm 0.020$~fm almost agrees with 
our previous value $r_{\rm skin}^{208}({\rm exp})=0.278 \pm 0.035$~fm of Ref.~\cite{Tagami:2020bee}.

\subsubsection{$^{58}$Ni}

\begin{figure}[H]
\begin{center}
 \includegraphics[width=0.452\textwidth,clip]{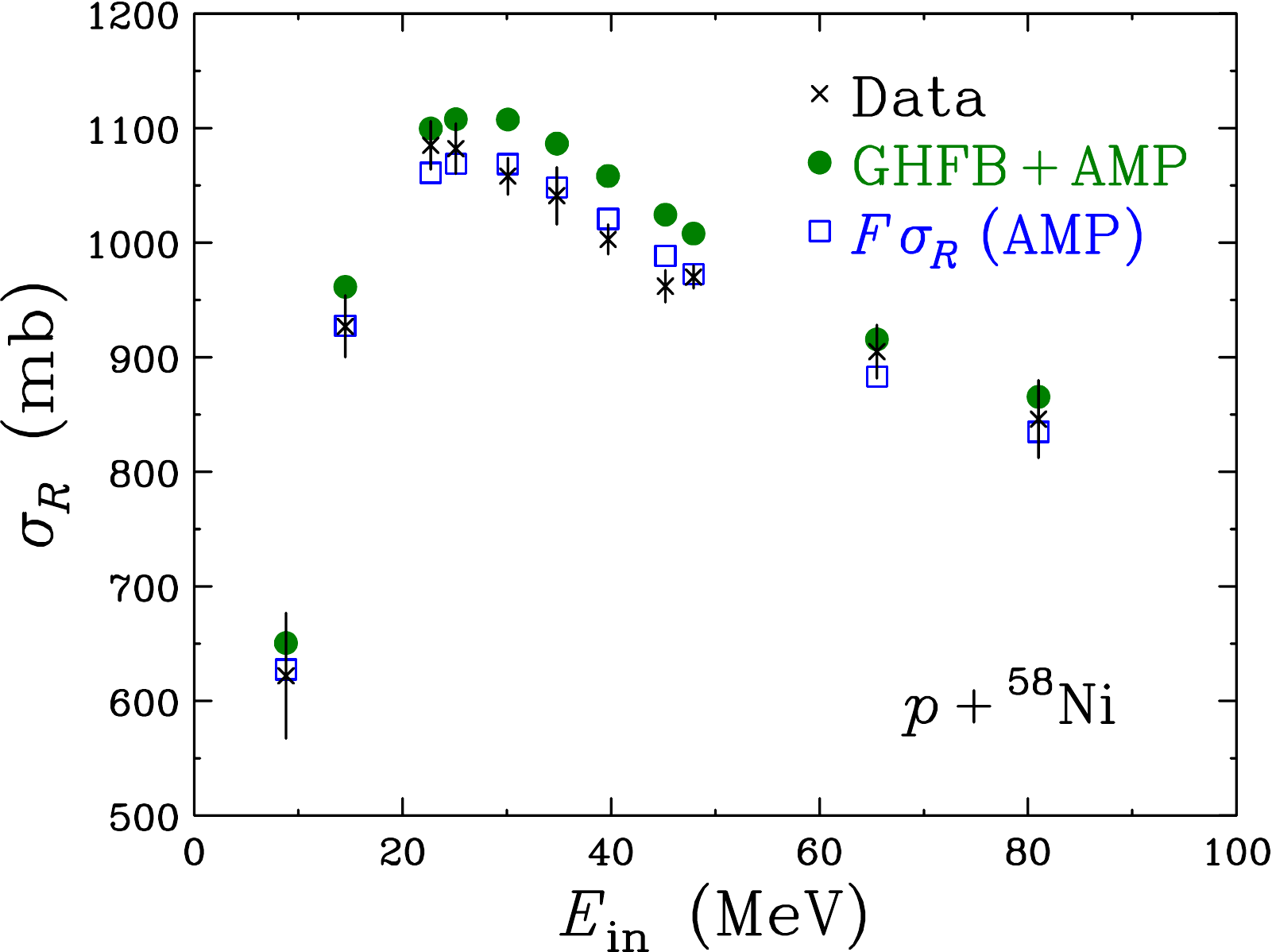}
 \caption{ 
 $E_{\rm in}$ dependence of reaction cross sections $\sigma_{\rm R}$ 
 for $p$+$^{58}$Ni scattering. 
 Closed circles denote results of the  D1S-GHFB+AMP densities. 
 Squares stand for $F \sigma_{\rm R}({\rm AMP})$ with $F=0.96473$.
 The data (crosses) are taken from 
 Refs.~\cite{Auce:2005ks,Ingemarsson:1999sra,EliyakutRoshko:1995fn,Dicello:1967zz,Bulman:1968ujl}.
   }
 \label{Fig-RXsec-p+Ni58}
\end{center}
\end{figure}

Figure~ \ref{Fig-RXsec-p+Ni58} shows $\sigma_{\rm R}$ 
as a function of $E_{\rm in}$ for $p$+$^{58}$Ni scattering.  
The results $\sigma_{\rm R}({\rm AMP})$ 
of the Kyushu $g$-matrix folding model with the D1S-GHFB+AMP densities (closed circles)  
almost reproduces data $\sigma_{\rm R}({\rm exp})$~\cite{Auce:2005ks,Ingemarsson:1999sra,EliyakutRoshko:1995fn,Dicello:1967zz,Bulman:1968ujl} in $10 \lsim E_{\rm in} \leq 81$MeV; 
note that the data has high accuracy of 2.7\%.

The result of ESP-F is $r_{m}({\rm exp})=3.711 \pm 0.010 $~fm. 
Using  the $r_{m}({\rm exp})$ and $r_{\rm p}({\rm exp})=3.685~{\rm fm}$~\cite{Angeli:2013epw}, 
we can obtain $r_{\rm skin}=0.055  \pm 0.010$~fm and $r_{\rm n}=3.740 \pm 0.010$~fm.  

A novel method for measuring nuclear reactions in inverse kinematics with stored ion beams was successfully used to extract the matter radius of $^{58}$Ni~\cite{Zamora:2017adt}. The experiment was performed at the experimental heavy-ion storage ring at the GSI facility. 
Their results determined from the differential cross section 
for $^{58}$Ni+$^{4}$He scattering are $r_m({\rm GSI})=3.70(7)$~fm, $r_p({\rm GSI})=3.68$~fm, 
$r_n({\rm GSI})=3.71(12)$~fm, $r_{\rm skin}({\rm GSI})=0.03(12)$~fm.

\subsubsection{$^{48}$Ca}

\begin{figure}[H]
\begin{center}
 \includegraphics[width=0.45\textwidth,clip]{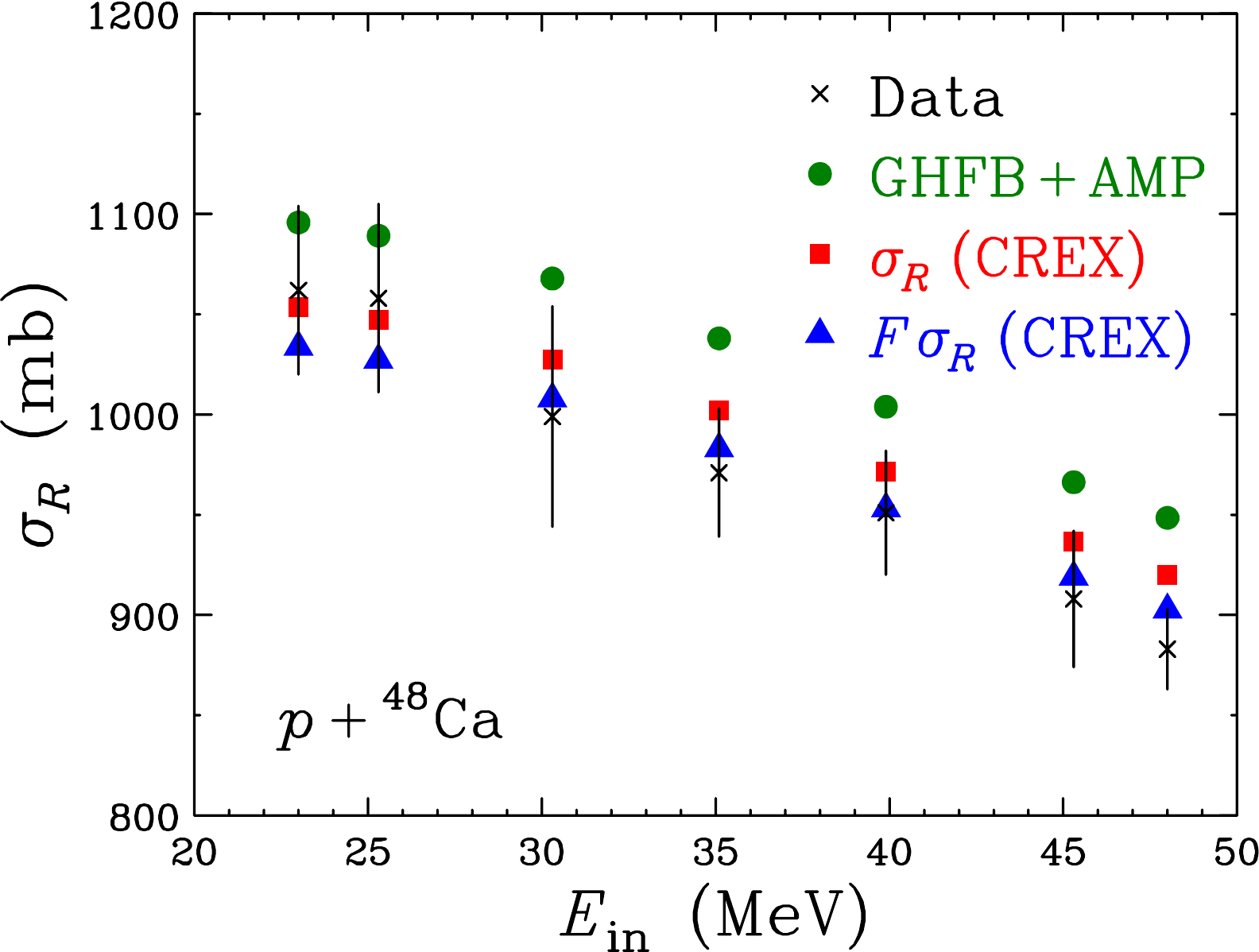}
 \caption{ 
 $E_{\rm in}$ dependence of reaction cross sections $\sigma_{\rm R}$ 
 for $p$+$^{48}$Ca scattering. 
 Circles denote results of the  D1S-GHFB+AMP densities, and 
 squares correspond to the results of the scaled densities based on $r_{\rm skin}^{48}({\rm CREX})$.  
 Triangles stand for $F \sigma_{\rm R}({\rm CREX})$ with $F=0.9810$.
  The data (crosses) are taken from Ref.~\cite{Carlson:1994fq}. 
   }
 \label{Fig-RXsec-p+Ca48}
\end{center}
\end{figure}

Figure~ \ref{Fig-RXsec-p+Ca48} shows  $\sigma_{\rm R}$ 
as a function of $E_{\rm in}$ for $p$+$^{48}$Ca scattering. 
The $\sigma_{\rm R}({\rm AMP})$  almost reproduce the data~\cite{Carlson:1994fq}. 
The results $\sigma_{\rm R}({\rm CREX})$ based on $r_{\rm n}({\rm CREX})$  
and $r_{\rm p}({\rm exp})$~\cite{Angeli:2013epw} are 
near the central values of the data~\cite{Carlson:1994fq}. 
$E_{\rm in}$ dependence of $\sigma_{\rm R}(E1{\rm pE})$ 
is similar to that of the data~\cite{Carlson:1994fq}. 
The $F \sigma_{\rm R}({\rm CREX})$ almost reproduce the central values of the  data.

The results of ESP-F are $r_{\rm skin}=0.103  \pm 0.022$~fm and $r_{\rm n}=3.488 \pm 0.022$~fm. 
Our skin value agrees with $r_{\rm skin}^{48}({\rm CREX})$. 

\subsubsection{$^{40}$Ca}

\begin{figure}[H]
\begin{center}
 \includegraphics[width=0.45\textwidth,clip]{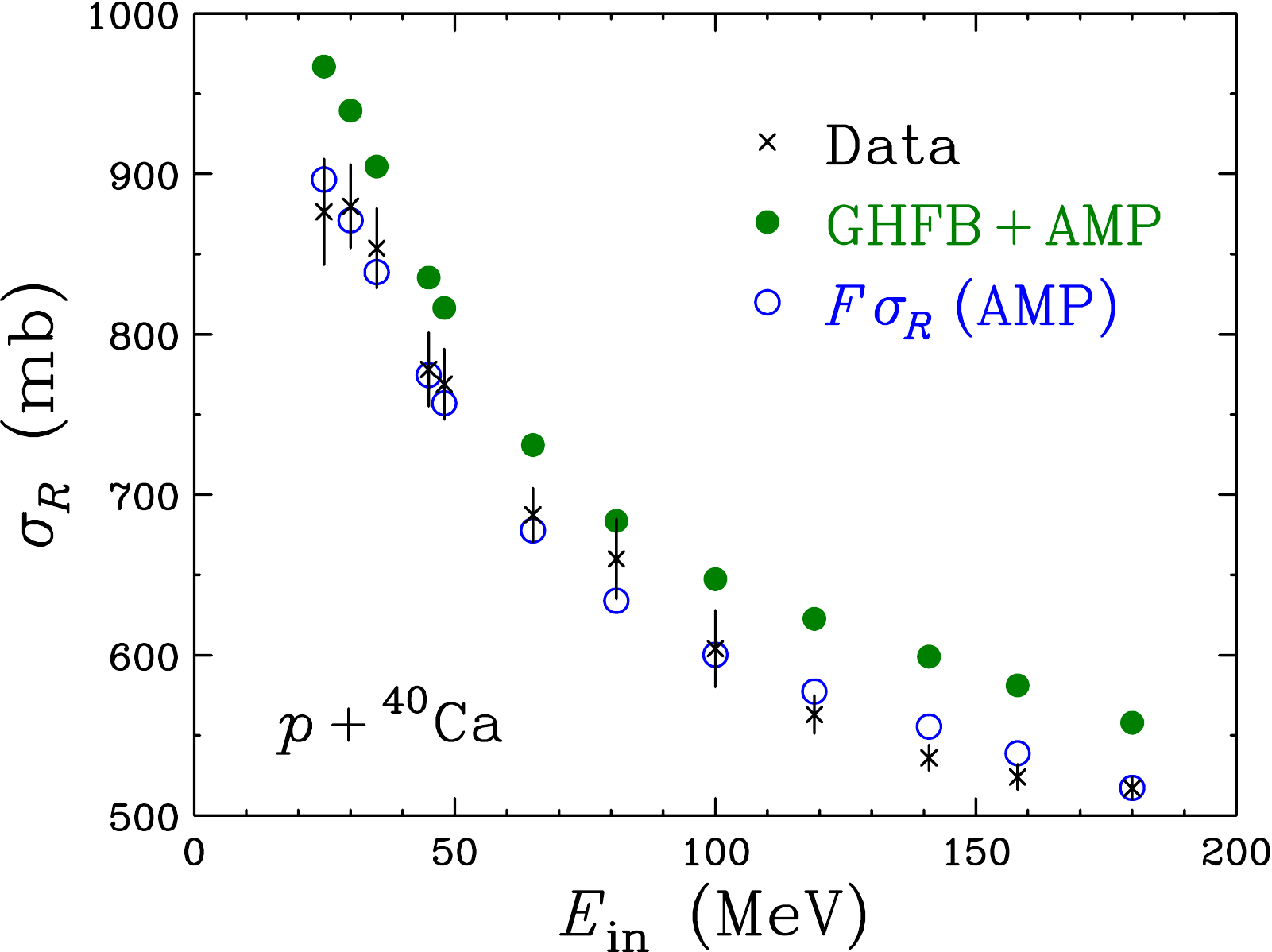}
 \caption{ 
 $E_{\rm in}$ dependence of reaction cross sections $\sigma_{\rm R}$ 
 for $p$+$^{40}$Ca scattering. 
 Closed circles denote results of the  original (D1S-GHFB+AMP) densities and the scaled ones, 
 Open circles correspond to   $F \sigma_{\rm R}({\rm AMP})$. 
 The data (crosses) are taken from 
 Refs.~\cite{Carlson:1975zz,Ingemarsson:1999sra,Auce:2005ks}. 
   }
 \label{Fig-RXsec-p+Ca40}
\end{center}
\end{figure}

Figure~ \ref{Fig-RXsec-p+Ca40} shows  $\sigma_{\rm R}$ 
as a function of $E_{\rm in}$ for $p$+$^{40}$Ca scattering. 
The Kyushu $g$-matrix folding model with the D1S-GHFB+AMP densities  
overestimates  $\sigma_{\rm R}({\rm exp})$~\cite{Carlson:1975zz,Ingemarsson:1999sra,Auce:2005ks}; 
note that the data has high accuracy of 2.7~\%. 
Note that $\sigma_{\rm R}({\rm AMP})=\sigma_{\rm R}({\rm PCNP})$, since $r_{m}({\rm AMP})$ is very close to 
$r_{m}({\rm RCNP})$.
$E_{\rm in}$ dependence of $\sigma_{\rm R}({\rm AMP})$ 
is similar to that of the data~\cite{Carlson:1975zz,Ingemarsson:1999sra,Auce:2005ks}.

The result od ESP-F is $r_{m}({\rm exp})=3.372 \pm 0.011 $~fm. 
Using  the $r_{m}({\rm exp})$ and $r_{\rm p}({\rm exp})=3.378~{\rm fm}$ of electron scattering, 
we can obtain $r_{\rm skin}({\rm exp})=-0.011  \pm 0.011$~fm and $r_{\rm n({\rm exp})}=3.367 \pm 0.011$~fm. 
Our results are close to those of shown in Ref.~\cite{Zenihiro:2018rmz}; see Table  \ref{reference values} 
for the values of Ref.~\cite{Zenihiro:2018rmz}. 

\subsubsection{$^{12}$C}

\begin{figure}[H]
\begin{center}
 \includegraphics[width=0.45\textwidth,clip]{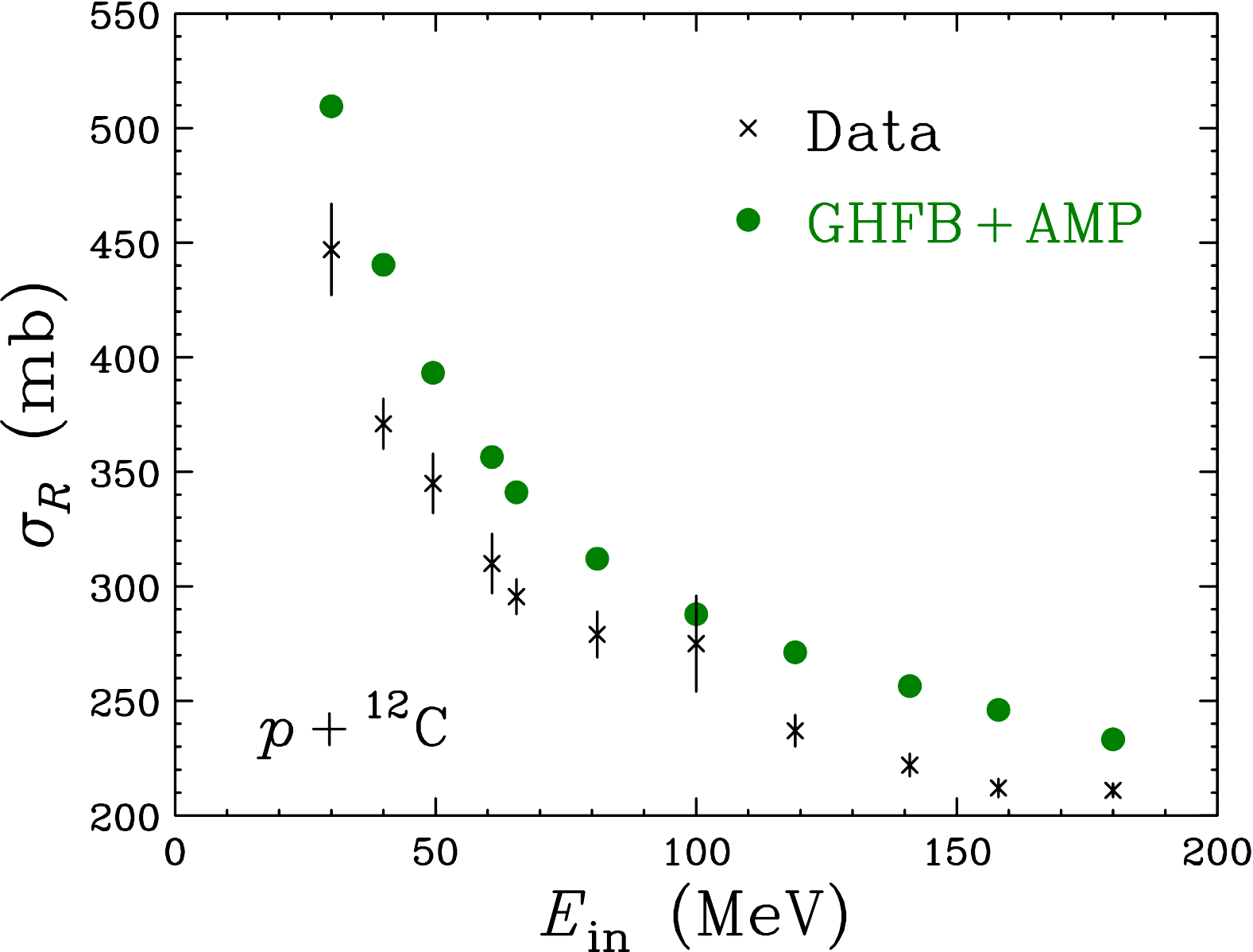}
 \caption{ 
 $E_{\rm in}$ dependence of reaction cross sections $\sigma_{\rm R}$ 
 for $p$+$^{12}$C scattering. 
 Closed circles denote results  $F \sigma_{\rm R}({\rm AMP})$. 
 The data (crosses) are taken from 
 Refs.~\cite{Auce:2005ks,Ingemarsson:1999sra,Menet:1971zz}. 
   }
 \label{Fig-RXsec-p+C12}
\end{center}
\end{figure}

Figure~ \ref{Fig-RXsec-p+C12} shows  $\sigma_{\rm R}$ 
as a function of $E_{\rm in}$ for $p$+$^{12}$C scattering. 
The results $\sigma_{\rm R}({\rm AMP})$ of D1S-GHFB+AMP overshoot data $\sigma_{\rm R}({\rm exp})$~\cite{Auce:2005ks,Ingemarsson:1999sra,Menet:1971zz}.
  
\begin{figure}[H]
\begin{center}
 \includegraphics[width=0.45\textwidth,clip]{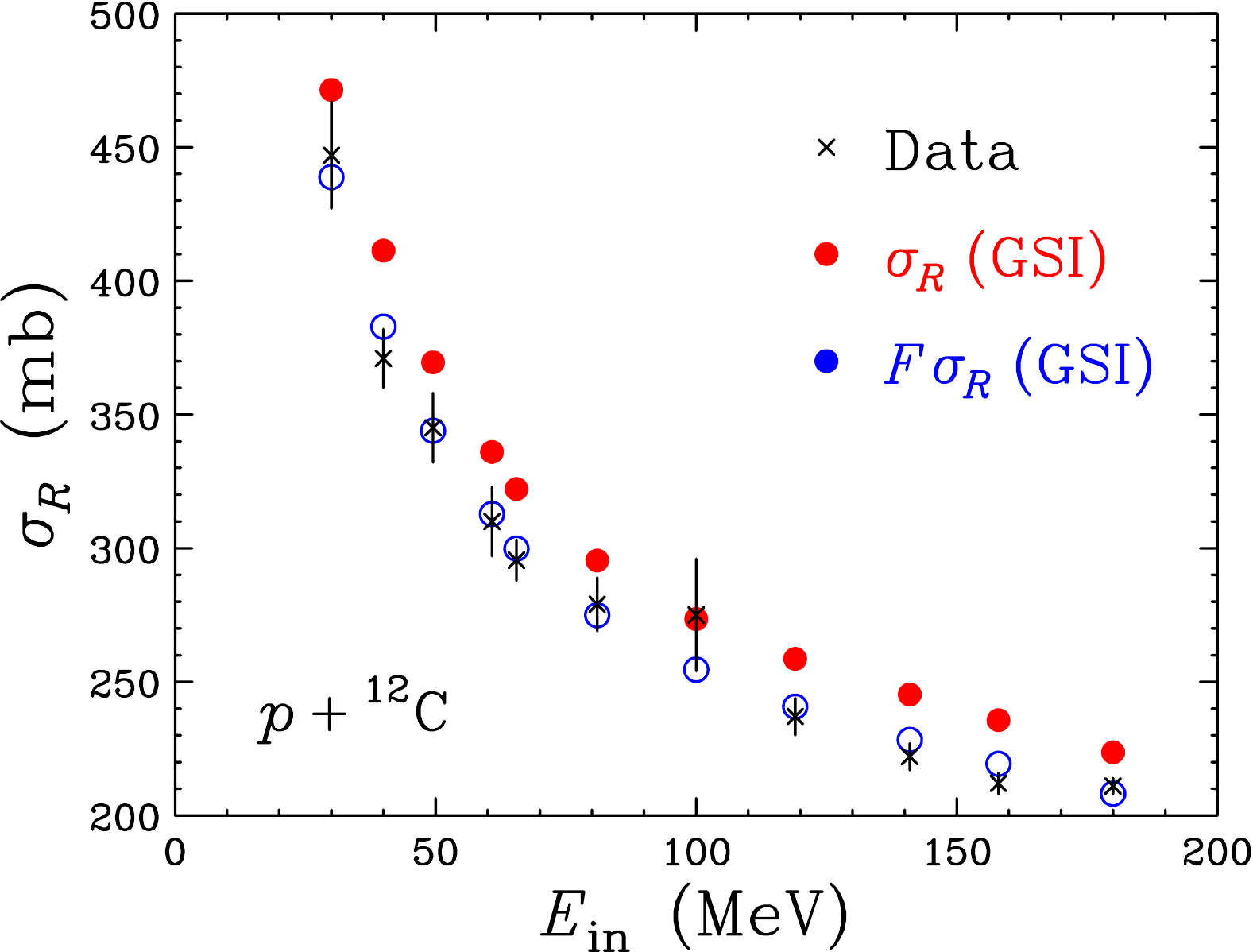}
 \caption{ 
 $E_{\rm in}$ dependence of reaction cross sections $\sigma_{\rm R}$ 
 for $p$+$^{12}$C scattering. 
 Closed circles denote results  $\sigma_{\rm R}({\rm GSI})$, while 
 open circles correspond  to $F \sigma_{\rm R}({\rm GSI})$. 
 The data (crosses) are taken from 
 Refs.~\cite{Auce:2005ks,Ingemarsson:1999sra,Menet:1971zz}. 
   }
 \label{Fig-RXsec-p+C12-2}
\end{center}
\end{figure}

Figure~ \ref{Fig-RXsec-p+C12-2} shows $\sigma_{\rm R}({\rm GSI})$ based on 
$r_{\rm m}({\rm GSI})$ and $r_{\rm p}({\exp})=2.327$~fm  of electron scattering for $p$+$^{12}$C scattering. 
The results $\sigma_{\rm R}({\rm GSI})$ are near the upper bound of 
$\sigma_{\rm R}({\rm exp})$~\cite{Auce:2005ks,Ingemarsson:1999sra,Menet:1971zz}. 
The $F \sigma_{\rm R}({\rm GSI})$ (open circles) are near the central values of 
$\sigma_{\rm R}({\rm exp})$. 

The result of ESP-F is $r_{m}({\rm exp})=2.340 \pm 0.009 $~fm. 
Using  the $r_{m}({\rm exp})$ and $r_{\rm p}({\rm exp})=2.327~{\rm fm}$, 
we can obtain $r_{\rm skin}({\rm exp})=0.026  \pm 0.009$~fm and $r_{\rm n}({\rm exp})=2.354  \pm 0.009$~fm.

Tanihata {\it et al.} determined $r_{m}$ from interaction cross sections 
for He, Li, Be, B isotopes~\cite{Tanihata:1988ub}. 
In Ref.~\cite{Ozawa:2001hb}, the experimental values on $r_{m}$ are accumulated  from $^{4}$He to $^{32}$Mg. 
Our result  $r_{m}({\rm exp})=2.340 \pm 0.009 $~fm is slightly smaller to $r_{m}({\rm GSI})=2.35(2)$~fm. 
As for neutron radius,  this is the case, because  $r_{n}({\rm exp})=2.354 \pm 0.009 $~fm and 
$r_{n}({\rm GSI})=2.37(2)$~fm.

\section{Summary}
\label{Summary}

In this paper, we consider the $^{208}$Pb, $^{58}$Ni, $^{40,48}$Ca, $^{12}$C, as stable nuclei and 
determine $r_{\rm skin}(\sigma_{\rm R})$, $r_{\rm m}(\sigma_{\rm R})$, $r_{\rm n}(\sigma_{\rm R})$ 
from measured $\sigma_{\rm R}$.
Our results on $r_{\rm skin}(\sigma_{\rm R})$, $r_{\rm m}(\sigma_{\rm R})$, $r_{\rm n}(\sigma_{\rm R})$, 
are summarized in Table \ref{TW values}. 
  Comparing Table \ref{TW values} 
 with Table  \ref{reference values}, we find that our results are close to the reference values.

\squeezetable
\begin{table}[htb]
\begin{center}
\caption
{Our results for  $r_{\rm m}$,  $r_{\rm n})$,  $r_{\rm skin}$. 
The radii are shown in units of fm.  
 }
\begin{tabular}{cccc}
\hline\hline
 & $r_{\rm m}(\sigma_{\rm R})$ &  $r_{\rm n}(\sigma_{\rm R})$ & $r_{\rm skin}(\sigma_{\rm R})$ \\
\hline
 $^{208}$Pb &  $5.627 \pm 	0.020 $ & $5.743  \pm 	0.020$ & $0.299  \pm 	0.020$ \\
 $^{58}$Ni &  $3.711  \pm	0.010 $ & $3.740  \pm	0.010 $ & $0.055  \pm	0.010 $ \\
 $^{48}$Ca &  $3.445 \pm	0.022$ & $3.488  \pm	0.022$ & $0.103  \pm	0.022$ \\
 $^{40}$Ca &  $3.372 \pm	0.011$ & $3.367 \pm	0.011$ & $-0.011 \pm	0.011$ \\
 $^{12}$C &  $2.340 	\pm 0.009 $ & $2.354 	\pm 0.009$ & $0.026 	\pm 0.009$ \\
\hline
\end{tabular}
 \label{TW values}
 \end{center} 
 \end{table}

\begin{figure}[H]
\begin{center}
 \includegraphics[width=0.45\textwidth,clip]{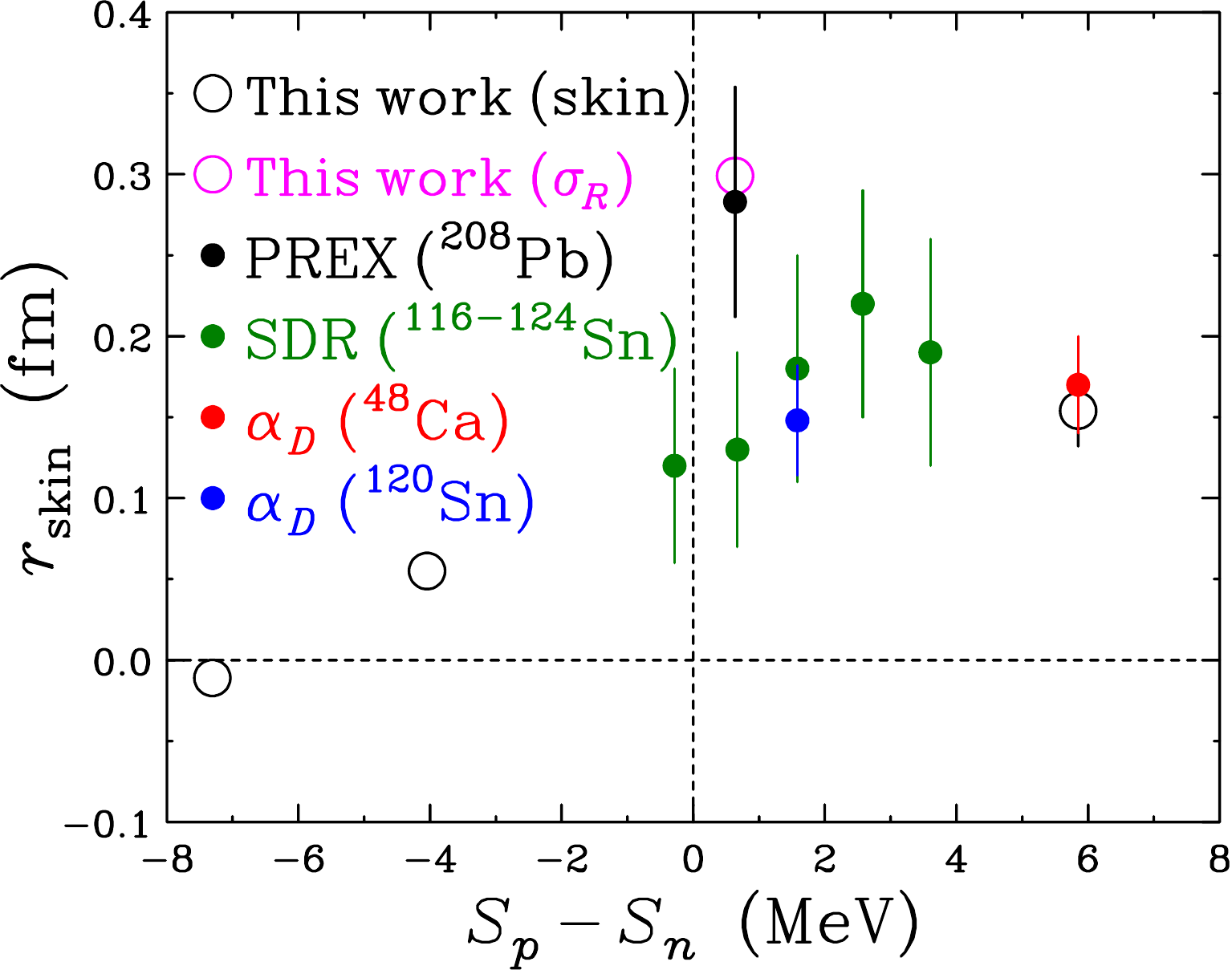}
 \caption{ 
 Skin values as a function of $S_{\rm p}-S_{\rm n}$. 
 The skin values determined from  measured $\sigma_{\rm R}$ are shown 
 with ``This work (skin)'' for $^{208}$Pb
 and with ``This work ($\sigma_{\rm R}$)'' for  $^{58}$Ni, $^{40,48}$Ca.
  The symbol `` $\a_D$'' means the results of the  $E1$ polarizability experiment ($E1$pE) for $^{120}$Sn~\cite{Hashimoto:2015ema} and 
$^{48}$Ca~\cite{Birkhan:2016qkr}. 
 The symbol ``PREX'' stands for the result deduced 
 from  $r_{\rm skin}^{208}({\rm PREX2}) = 0.283\pm 0.071{\rm fm}$. 
 Open circles stand for the results of this work. 
The symbol ``SDR" shows the results~\cite{Krasznahorkay:1999zz} of the measurement based on the isovector spin-dipole resonances (SDR) in Sn isotopes. 
  The data ( closed circles with error bar) are taken from 
 Refs.~\cite{Krasznahorkay:1999zz,Hashimoto:2015ema,Birkhan:2016qkr,Adhikari:2021phr}.
}
 \label{Fig-skins}
\end{center}
\end{figure}

\begin{figure}[H]
\begin{center}
 \includegraphics[width=0.45\textwidth,clip]{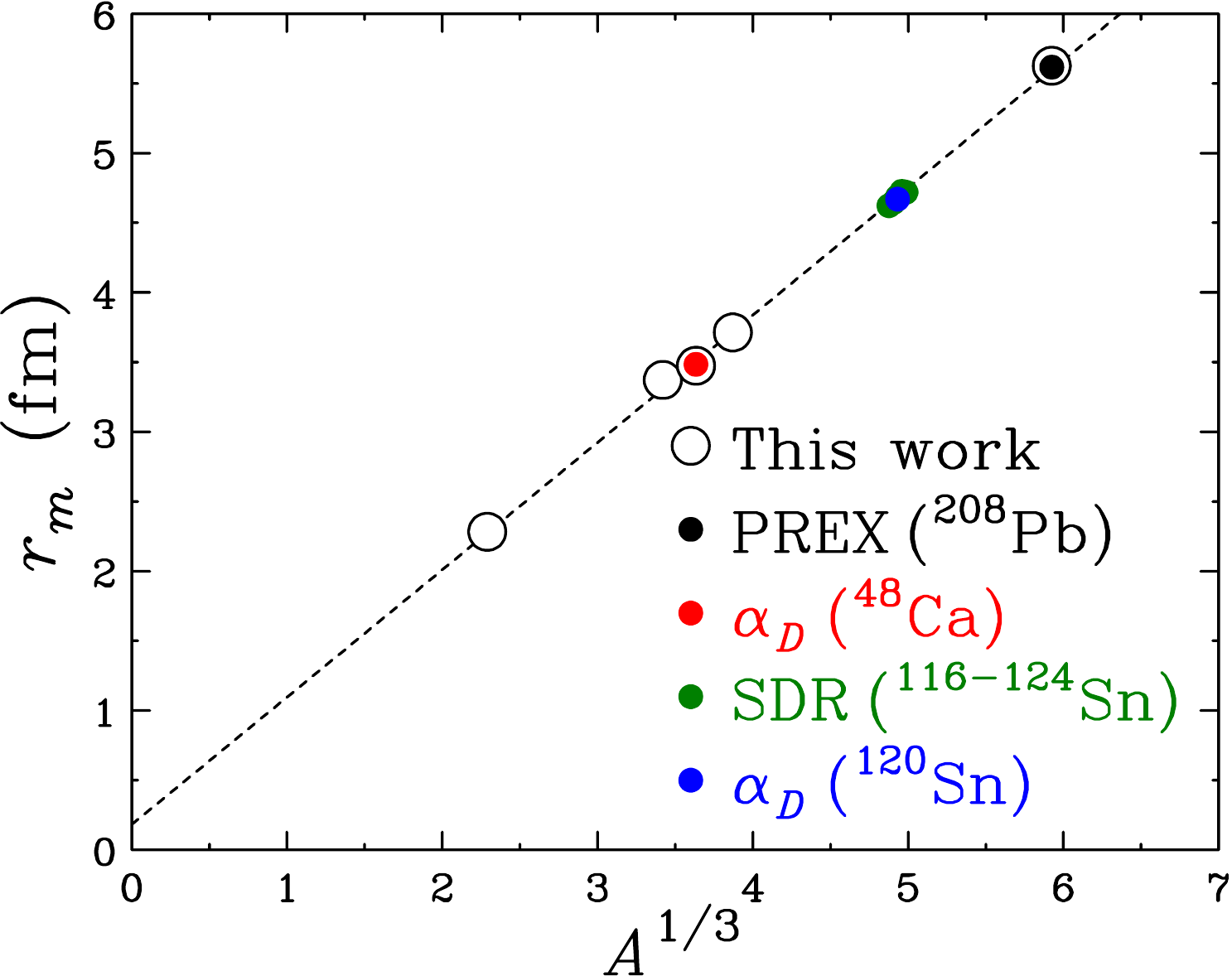}
 \caption{ 
 Matter radii $r_{\rm m}$ as a function of mass number $A^{1/3}$. 
 The symbol `` $\a_D$'' means the results of the 
 $E1$ polarizability experiment ($E1$pE) for $^{120}$Sn~\cite{Hashimoto:2015ema} and 
$^{48}$Ca~\cite{Birkhan:2016qkr}. 
 The symbol ``PREX'' stands for the result deuced from 
  $r_{\rm skin}^{208}({\rm PREX2}) = 0.283\pm 0.071{\rm fm}$. 
The symbol ``SDR" shows the results~\cite{Krasznahorkay:1999zz} of the measurement based on the isovector spin-dipole resonances (SDR) in the Sb isotopes. 
 Open circles stand for the results of this work. 
 The dashed line is a guide for the eyes.
  The data ( closed circles with error bar) are taken from 
 Refs.~\cite{Krasznahorkay:1999zz,Hashimoto:2015ema,Birkhan:2016qkr,Adhikari:2021phr}.
}
 \label{Fig-Rm}
\end{center}
\end{figure}

\begin{figure}[H]
\begin{center}
 \includegraphics[width=0.45\textwidth,clip]{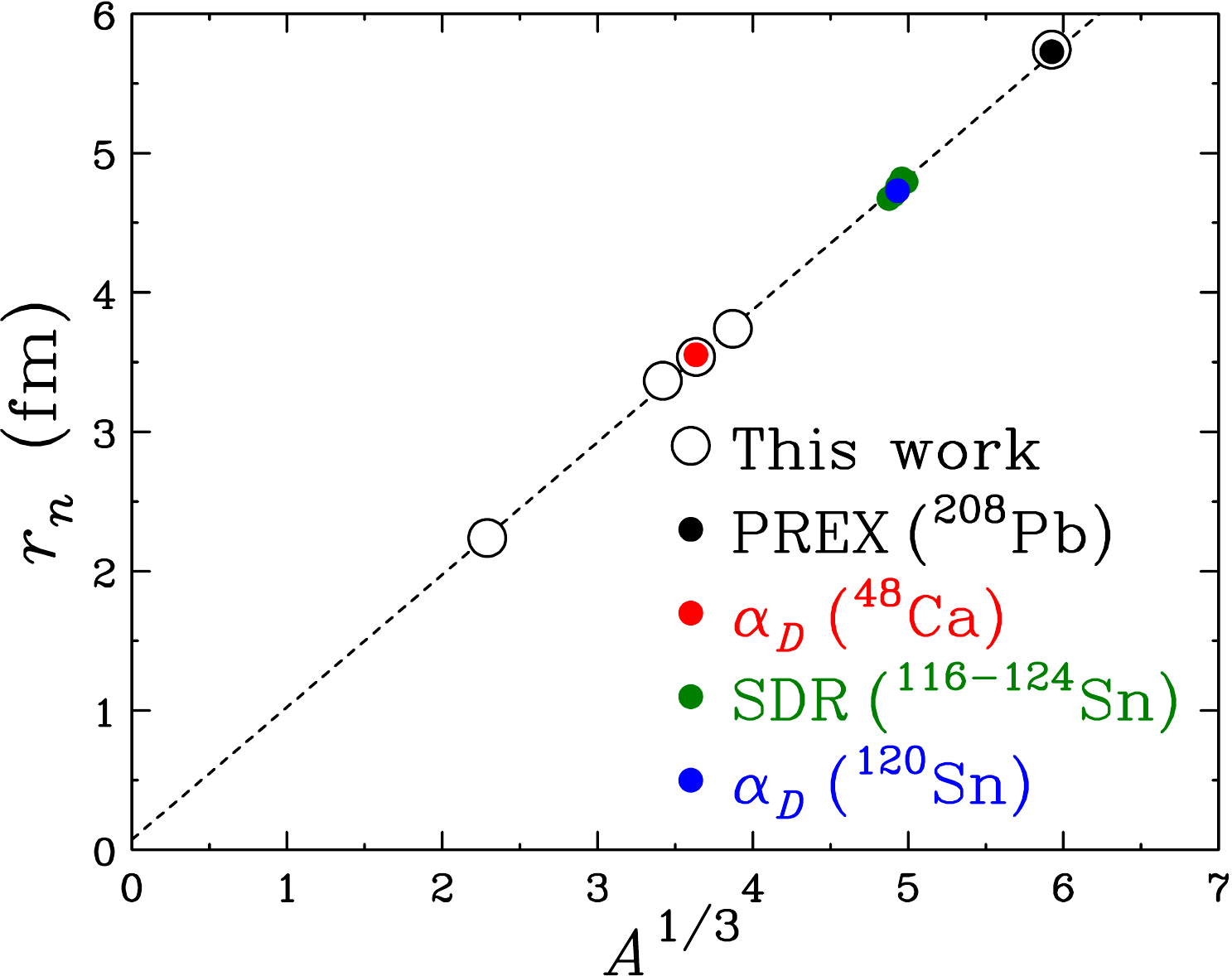}
 \caption{ 
 Neutron radii $r_{\rm n}$ as a function of mass number $A^{1/3}$. 
 The symbol `` $\a_D$'' means the results of the 
 $E1$ polarizability experiment ($E1$pE) for $^{120}$Sn~\cite{Hashimoto:2015ema} and 
$^{48}$Ca~\cite{Birkhan:2016qkr}. 
 The symbol ``PREX'' stands for the result deuced 
 from  $r_{\rm skin}^{208}({\rm PREX2}) = 0.283\pm 0.071{\rm fm}$. 
The symbol ``SDR" shows the results~\cite{Krasznahorkay:1999zz} of the measurement based on the isovector spin-dipole resonances (SDR) in the Sb isotopes. 
 Open circles stand for the results of this work. 
 The dashed line is a guide for the eyes. 
   The data ( closed circles with error bar) are taken from 
 Refs.~\cite{Krasznahorkay:1999zz,Hashimoto:2015ema,Birkhan:2016qkr,Adhikari:2021phr}.
}
 \label{Fig-Rn}
\end{center}
\end{figure}

We show mass-number ($A$) dependence of stable nuclei in Figs.~\ref{Fig-skins},~\ref{Fig-Rm}, \ref{Fig-Rn}. 
Figure \ref{Fig-skins} shows skin values as a function of $S_{\rm p}-S_{\rm n}$, 
where $S_{\rm p}$ ($S_{\rm n}$) is the proton (neutron) separation energy. 
The skin values  $r_{\rm skin}(\sigma_{\rm R})$ determined from measured $\sigma_{\rm R}$ 
for $^{208}$Pb, $^{58}$Ni, $^{40,48}$Ca are compared with the data of PREX2~\cite{Adhikari:2021phr}, 
$^{116,118,120,122,124}$Sn~\cite{Krasznahorkay:1999zz,Hashimoto:2015ema}, 
$^{48}$Ca~\cite{Birkhan:2016qkr}.
Our results are consistent with the previous experimental skin-values. 

Figure \ref{Fig-Rm} shows matter radii $r_{\rm m}$ as a function of $A^{1/3}$.
For $^{208}$Pb, $^{116,118,120,122,124}$Sn, $^{48}$Ca, 
the $r_{\rm m}$ are derived from the corresponding skin values~\cite{Adhikari:2021phr,Krasznahorkay:1999zz,Hashimoto:2015ema,Birkhan:2016qkr} and the corresponding $r_p$ of electron scattering. 
For $^{12}$C, $^{40,48}$Ca, $^{58}$Ni, $^{208}$Pb, our results are added. 
Our results are consistent with the previous works.

Figure \ref{Fig-Rn} shows neutron radii $r_{\rm n}$ as a function of mass number $A^{1/3}$. 
For $^{208}$Pb, $^{116,118,120,122,124}$Sn, $^{48}$Ca, 
the $r_{\rm n}$ are derived from the corresponding skin values~\cite{Krasznahorkay:1999zz,Hashimoto:2015ema,Birkhan:2016qkr,Adhikari:2021phr} and the corresponding $r_p$ of electron scattering. 
For $^{12}$C, $^{40,48}$Ca, $^{58}$Ni, $^{208}$Pb, our results are added. 
Our results are consistent with the previous works.

\noindent
\begin{acknowledgments}
We would like to thank Dr. Toyokawa for providing his code and Prof. M. Nakano for useful information. 
\end{acknowledgments}



\bibliography{Folding-v15}

\end{document}